\documentclass[12pt]{iopart}

\usepackage{graphicx}
\usepackage{dcolumn}
\usepackage{bm}

\begin{document}

\title[Redox functionality mediated by adsorbed oxygen on a Pd-oxide film]
{Redox functionality mediated by adsorbed oxygen on 
a Pd-oxide film over a Pd(100) thin structure: \\
A first-principles study}

\author{$^1$K Kusakabe, $^2$K Harada, $^1$Y Ikuno, 
and $^1$H Nagara}

\address{
$^1$Graduate School of Engineering Science, Osaka University, 
1-3 Machikaneyama-cho, Toyonaka, Osaka 560-8531, Japan}
\ead{
\mailto{kabe@mp.es.osaka-u.ac.jp}
}

\address{
$^2$Institute for Solid State Physics, University of Tokyo, 
Kashiwanoha, Kashiwa, Chiba 277-8581, Japan}

\begin{abstract}
Stable oxygen sites on a PdO film over a Pd(100) thin structures 
with a ($\sqrt{5}\times\sqrt{5}$)$R27^\circ$ surface-unit cell 
are determined using the first-principles electronic structure calculations 
with the generalized gradient approximation. 
The adsorbed monatomic oxygen goes to a site bridging 
two 2-fold-coordinated Pd atoms or to a site 
bridging a 2-fold-coordinated Pd atom and a 4-fold-coordinated Pd atom. 
Estimated reaction energies of 
CO oxidation by reduction of the oxidized PdO film and 
N$_2$O reduction mediated by oxidation of the PdO film 
are exothermic. 
Motion of the adsorbed oxygen atom between the two stable sites 
is evaluated using the nudged elastic band method, 
where an energy barrier for a translational motion of 
the adsorbed oxygen may become $\sim 0.45$\,eV, 
which is low enough to allow fluxionality of the surface oxygen 
at high temperatures. 
The oxygen fluxionality is allowed by existence of 
2-fold-coordinated Pd atoms on the PdO film, 
whose local structure has similarity to that of Pd catalysts 
for the Suzuki-Miyaura cross coupling. 
Although NO$_x$ 
(including NO$_2$ and NO) reduction is not always catalyzed 
only by the PdO film, 
we conclude that there may happen continual redox reactions 
mediated by oxygen-adsorbed PdO films 
over a Pd surface structure, when the influx of NO$_x$ and CO continues, 
and when the reaction cycle is kept on a well-designed oxygen surface. 
\end{abstract}


\section{\label{Introduction}Introduction}

Oxidation and reduction processes on palladium surfaces are 
attracting interest, since the understanding of catalytic reactions 
on palladium is a key point to explore functionality of this solid catalyst 
working as the three-way catalyst for the automotive 
emissions control.\cite{Catalyst} 
When oxidation of Pd proceeds, 
various stable Pd-oxide films are created to cover 
the Pd surface depending on the surface morphology.\cite{Mittendorfer} 
A stable surface structure of an oxidized Pd(100) surface is known to be 
($\sqrt{5}\times\sqrt{5}$)$R27^\circ$ PdO mono-layer.\cite{Todorova,Kostelnk} 
Formation of a PdO thin film stabilizes 
the Pd(100) surface against further oxidation. 
Interestingly, catalytic reactivity is often attributed to surface 
oxides.\cite{Klikovitz,Westerstrom} 
Here, two questions arise. Is there a process to reach oxidation 
of the Pd substrate below the PdO film? 
How high is the reactivity of the PdO film 
itself in a catalytic reaction process? 

In a study of CO oxidation at Pd(100), 
Rogal, Reuter and Scheffler have shown 
surface phase diagram in constrained thermodynamic equilibrium 
obtained using the first-principles statistical mechanics.\cite{Rogal} 
They carefully constructed the phase diagram with 
converged slab structures with surfaces reacted with oxygen and/or CO. 
Interestingly, no coadsorption structures on Pd(100) in equilibrium 
was found. But, Langmuir-Hinshelwood reaction processes 
were expected around the phase boundary between surface oxide phases 
and CO adsorbed surfaces. 
In their surface phase diagram, we can see a wide stable 
phase of ($\sqrt{5}\times\sqrt{5}$)$R27^\circ$ surface oxide structure, 
which was found in the experiments. 

In consideration of the three-way catalysts, 
CO oxidation happens continually in a reactor by a gas flow. 
Thus, one needs to consider reactivity of the catalyst 
in a low partial pressure of CO. 
A key to understand total reactivity in this condition 
may be found in a point that, 
on a three-way catalyst material, 
other surface structures coexist with a Pd surface. 
The phase diagram of Rogal suggests that surface oxide formation 
without any extra adsorbents on the Pd oxide film is favored 
in a wide pressure range. 
Here, we may start from re-considering reactivity of 
surface oxides in a non-equilibrium condition 
with fluctuating density of surface oxygen. 
There, adsorbed oxygen atoms might come from 
a catalytic material structure other than the PdO film. 
Since it is a hard task to treating all the possibility, 
we just consider a single adsorbed oxygen atom assumed to be 
provided by other surfaces of the catalytic material 
than the PdO film 
and test its stability and reactivity. 
Fluctuating behavior of the surface oxygen will be called fluxionality. 
For this possible motion of oxygen atoms, 
morphological softness is preferable. 
If this character is expected in nano-meter-scale Pd structures, 
the condition to have fluxionality may be relaxed. 
Thus, we should also search for a simulation result 
which might open our eyes. 

To study reactivity of the ($\sqrt{5}\times\sqrt{5}$)$R27^\circ$ PdO 
film on a Pd(100) surface, 
we performed the first-principles structural optimization simulations 
to obtain stable oxygen-adsorbed structures of the PdO thin film. 
Slab models of PdO/Pd(100) were used for the simulation and 
thus the calculation data should be interpreted as results for 
a thin Pd structure with an oxidized surface. 
Two stable sites were found in this study. 
Stability of these sites depends on local atomic configuration 
of the whole Pd-oxide structure in the simulation. 
Choice of the structure of the thin Pd substrate even relaxes 
conditions for possible oxygen migration. 
We evaluated reaction energies of 
reduction processes for NO$_x$ (NO, NO$_2$, or N$_2$O) 
and a CO oxidation process to estimate functionality 
of the PdO surface as a redox catalyst. 
Characteristic features of the adsorbed reactive oxygen atom 
are explored by analyzing the local electronic density of states and 
barrier heights of oxygen migration. 
The estimated transition path of the oxygen atom 
reveals that there can exist fluxionality 
analogous to motions of 
fluxional function groups in molecules\cite{Fluxionality} and clusters. 
Finally, we will summarize our conclusions, in which 
an expected mechanism of the catalytic function is proposed. 

\section{\label{Methods}Methods of calculations}

A generalized gradient approximation\cite{PBE,PBE2} based on the 
density functional theory\cite{Hohenberg-Kohn,Kohn-Sham} 
was adopted in our calculation. 
The calculation was done using the plane-wave expansion with 
the ultra-soft pseudo potentials\cite{Vanderbilt} 
realized in the Quantum-espresso 
ver 3.2.3 package.\cite{Baroni} 
Conditions for the simulation were the followings. 
Cut-off energies for the wave function and the charge density 
were 408\,eV and 2721\,eV, respectively. 
The $k$-point sampling was done with $8\times 8\times 1$ 
mesh points in the first Brillouin zone. 

As a typical model structure, we consider a slab model 
which consist of a PdO film on two Pd layers. 
We call this ``the model I'', or ``the thin model''. 
This model is very thin. 
We will show that some characteristic properties of the PdO film 
on this thin Pd layers are qualitatively different from 
those of the PdO film on the stable Pd (100) surface, 
although its structure is only slightly different from that of 
the PdO on the stable Pd (100) surfaces.
The structure of the model I 
gives us an interesting result on the oxygen fluxionality. 
We compare the result with another structure with a PdO film 
on up tp five Pd sublayers, 
which we call ``the model II'' or ``the thick model''. 
The thickness of Pd layers of the model II is enough to reproduce 
the PdO film structure on Pd(100) surface, since 
we have the converged PdO structure for 3, 4, or 5 Pd layers as substrates.

To check validity of the simulation for the model I, 
we performed several test simulations. 
A repeated slab model with a vacuum layer of 
7 \AA \, thickness was employed for each simulation. 
Convergence is seen as follows: 
When thickness of the vacuum layer is increased from 7\AA 
to 15\AA, 
atomic positions changed slightly only with absolute error of 
less than $10^{-4}$ \AA. 
The total energy changes only 
less than $7\times 10^{-3}$ eV for the model I. 

For the structural optimization of the model II, 
we prepared the PdO structure of a surface unit cell of
($\sqrt{5}\times\sqrt{5}$)$R27^\circ$ above 
3, 4, or 5 atomic layers of Pd, 
which is a structure with up to five Pd sublayers. 
At the start of the simulations, the size of the surface unit cell 
of the PdO was set so that the lattice constants of the Pd sublattice 
is equal to the lattice constants of an equilibrium lattice of the bulk Pd 
and kept fixed in each optimization calculations. 
All of the relative atomic positions are relaxed in each simulations. 
The optimized structures reproduce the former result\cite{Kostelnk} 
qualitatively and almost quantitatively 
(See Table\,\ref{Str_param} and also discussion in the final paragraph of 
the section\,\ref{Surface_migration}). 

However, when we look at the model I, 
the PdO film showed a little modified structure from that of the model II. 
For the comparison, we summarize the optimized 
structural parameters of the models I and II in Table \ref{Str_param}. 
We note here that the distance between Pd layers are:
$d_{{\rm PdO}-{\rm Pd}_2}=2.6773\AA$ and 
$d_{{\rm Pd}_2-{\rm Pd}_3}=2.0869\AA$ for the model I 
and 
$d_{{\rm PdO}-{\rm Pd}_2}=2.5464\AA$,
$d_{{\rm Pd}_2-{\rm Pd}_3}=2.0529\AA$,
$d_{{\rm Pd}_3-{\rm Pd}_4}=2.0526\AA$,
$d_{{\rm Pd}_4-{\rm Pd}_5}=2.0517\AA$, and 
$d_{{\rm Pd}_5-{\rm Pd}_6}=2.0226\AA$ for the mode II, 
where ${\rm Pd}_i$ with $i=2,3,\cdots, 6$ represents 
the $i$th Pd sublayer which consists up to 5 substrate layers. 

\begin{figure}
\begin{center}
\includegraphics[height=10cm]{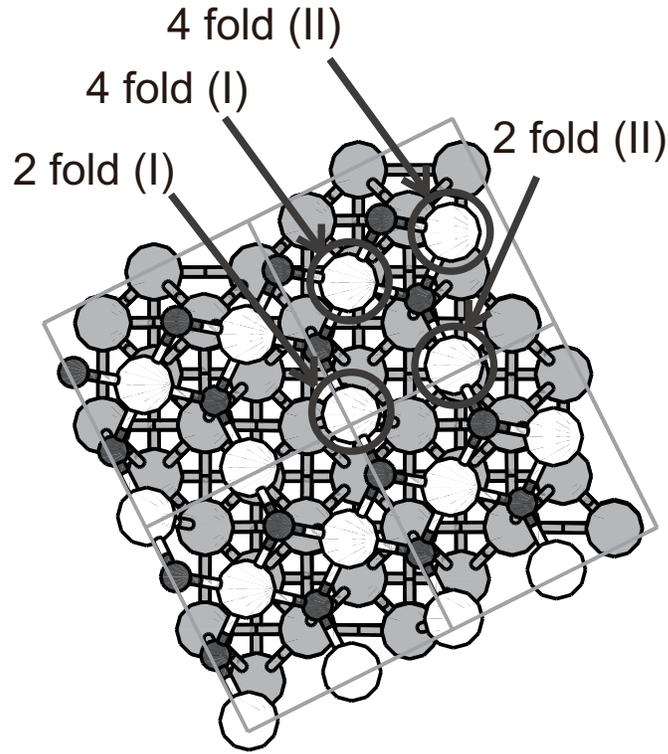}
\end{center}
\caption{The optimized ($\sqrt{5}\times\sqrt{5}$)$R27^\circ$ 
PdO film on Pd(100) surface of the model I 
obtained by the present simulation. 
White spheres are Pd atoms in the PdO film, 
bright gray spheres are Pd atoms in the substrate, 
and dark gray spheres are oxygen atoms. 
Each square represents a unit cell. 
In a unit cell, there are 
two 4-fold-coordinated sites and two 2-fold-coordinated sites, 
of Pd atoms, which are marked by circles. 
\label{PdO-structure-I}}
\end{figure}

The structure shown in Figure \ref{PdO-structure-I} 
is an optimized structure of the model I, which 
was obtained by the structural optimization, 
in which the inter atomic forces were reduced to 
less than $1.3 \times 10^{-5}$[eV/a.u.]. 
In the optimized film structure, 
4-fold-coordinated sites and 2-fold-coordinated sites of Pd exist 
as depicted in Figure \ref{PdO-structure-I}. 
Existence of the 2-fold-coordinated Pd atoms 
is important for our discussion. 

\begin{table}
\begin{center}
\begin{tabular}{lrrr}
\hline 
 & $x$(\AA) & $y$(\AA) & $\Delta z$(\AA) \\
\hline
Pd &       3.916 &   2.976 &  -0.052 \\
Pd &       1.020 &   5.182 &   0.238 \\
Pd &      -1.720 &   3.882 &   0.198 \\
Pd &       1.346 &   1.448 &  -0.384 \\
O  &       0.248 &   3.292 &  -0.369 \\
O  &       2.976 &   4.774 &  -0.473 \\
O  &      -0.846 &   5.560 &   1.066 \\
O  &       1.885 &   6.916 &   1.012 \\
\hline
Pd  &      3.932 &  3.046 &  -0.040 \\
Pd  &      1.120 &  5.192 &   0.100 \\
Pd  &     -1.642 &  3.840 &   0.095 \\
Pd  &      1.195 &  1.618 &  -0.155 \\
O   &      0.275 &  3.403 &  -0.531 \\
O   &      3.042 &  4.808 &  -0.533 \\
O   &     -0.764 &  5.548 &   0.761 \\
O   &      1.965 &  6.924 &   0.781 \\
\hline
\end{tabular}
\end{center}
\caption{
\label{Str_param}
Optimized structural parameters of the PdO film in the model I 
[two Pd layers below the PdO film] 
(upper half of the table) and the model II 
[four Pd layers below the PdO film] (lower half of the table). 
The coordinates $x$ and $y$ are in-plane coordinates, 
while $\Delta z$ gives shift in a vertical structural parameter 
from the center of $z$ coordinates of Pd atoms in the PdO layer.
}
\end{table}

\section{\label{Structure}Optimized structures of oxygen adsorbed PdO films.}

Each simulation for the 
determination of the oxygen-adsorbed site 
was started from the optimized structure of 
the PdO thin film shown in the last section. 
At first, we prepared four different initial structures using the model I. 
A single atomic oxygen was put on a Pd atom in the topmost 
oxidized layer and then the position of the oxygen 
as well as the surface atomic structure was optimized. 
Note that we have four distinct Pd sites in the unit cell of 
the ($\sqrt{5}\times\sqrt{5}$)$R27^\circ$ structure. 
We performed four simulations to optimize the oxygen adsorbed 
film structure, and we obtained two stable structures. 

\begin{figure}
\begin{center}
\includegraphics[height=10cm]{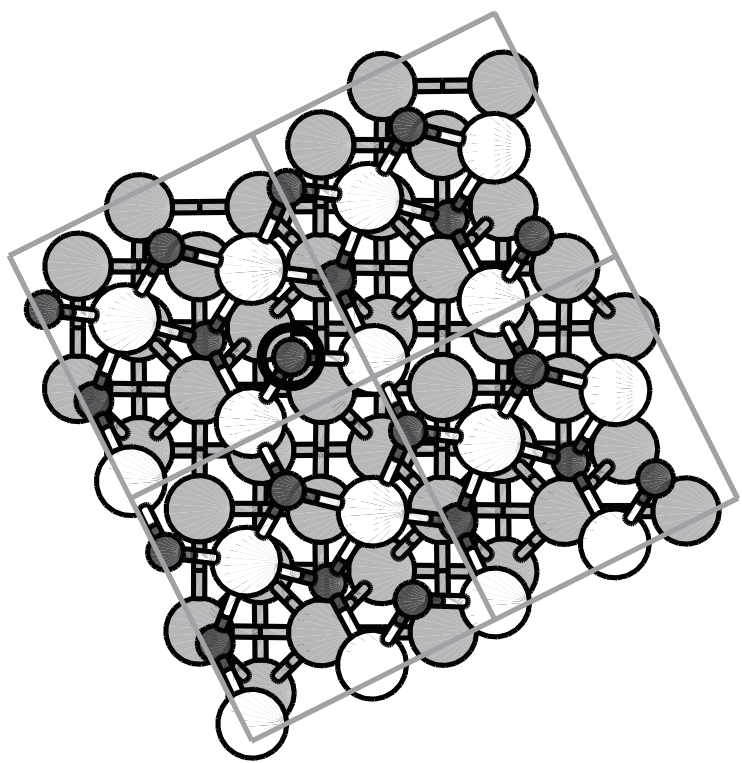}
\caption{The atomic configuration of the bridge-site (I) structure.
The adsorbed oxygen (O1) shown by a circle connects two 
2-fold-coordinated Pd atoms. 
\label{Bridge-structure-I}}
\end{center}
\end{figure}

\begin{figure}
\begin{center}
\includegraphics[height=10cm]{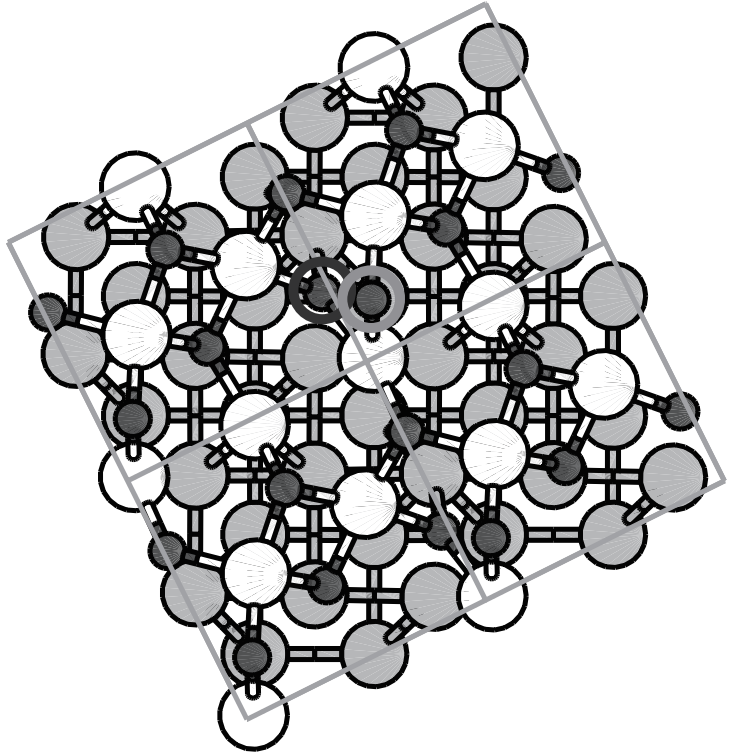}
\caption{The atomic configuration of the bridge-site (II) structure.
The adsorbed oxygen shown by a bright circle 
connects a 2-fold-coordinated Pd atom and a 4-fold-coordinated Pd atom.
Another oxygen atom, which formed the PdO film before 
the optimization, and is shown by a dark circle, 
has new bond connections with Pd atoms in the second Pd layer.
\label{Hollow-structure-I}}
\end{center}
\end{figure}

The obtained structures for adsorption of an oxygen atom per 
a surface unit cell of ($\sqrt{5}\times\sqrt{5}$)$R27^\circ$ PdO 
are shown in Figure \ref{Bridge-structure-I} and \ref{Hollow-structure-I}. 
The first one has an oxygen atom at 
a bridge site (Figure \ref{Bridge-structure-I}), 
in which the adsorbed oxygen 
bridges two Pd atoms in the 2-fold coordination. 
The other has an oxygen atom at 
another bridge site (Figure \ref{Hollow-structure-I}), 
in which the adsorbed oxygen atom has bond connections 
with a Pd atom in the 2-fold coordination and another Pd atom 
in the 4-fold coordination. 
We call the former site the bridge site (I) and 
the latter site the bridge site (II). 
In the bridge site (II) structure, 
a surface oxygen atom which had a bond connection 
with the two Pd atoms changes its bond connections: 
One of the bond connections is cut through the oxygen adsorption, 
and a new bond with a Pd atom in the second Pd layer is created. 
Thus, the subsurface oxidation happens in this process of 
the monoatomic oxygen adsorption. 

To confirm the subsurface oxidation, 
we evaluated the L\"{o}wdin charges. 
A value on a Pd atom in the second layer 
changes from 9.8995 to 9.7614, when 
the position of the adsorbed oxygen 
changes from the bridge site (I) to the bridge site (II). 
The value of $\sim$ 9.75 is obtained, when the Pd atom 
has a bond connection with oxygen in a PdO film. 
Thus we can say that the subsurface oxidation proceeds 
through the formation of the bridge site (II) structure in the model I. 

We now compare the total energy of the two solutions. 
The former bridge site (I) 
is energetically favorable by 52.2\,meV 
compared to the latter bridge site (II). 
The bridge site (I) is a natural structure, since the adsorbed oxygen 
atom forms bond connections with two 2-fold-coordinated Pd atoms. 
In a unit cell, there are two bridge sites (I). 
As discussed later, we will find a migration path 
of an oxygen atom going from a bridge site (I) to the other 
bridge site (I). 

The bridge site (II) may be regarded as an intermediate step 
of the subsurface oxidation. 
A surface oxygen atom originally forming 
a bond in the PdO top layer pproaches to 
a Pd atom in the second Pd layer, forming a new bond connection with it. 
The adsorption energy 
is a little larger for the bridge site (I) 
than that of the bridge site (II). 
This result also suggests 
that the reaction of subsurface oxidation does not easily occur. 

Next, we tried to find these two bridge sites 
using the model II. 
In this thick model, however, we could not find the bridge site (II). 
Even if we started the optimization of the adsorbed oxigen 
located approximately at a position of the bridge site (II), 
it went to the bridge site (I). 
Another simulation searching an oxigen-migration path from 
a bridge site (I) to another bridge site (I) did not 
show any structure similar to the bridge site (II) structure 
in the path. 
Thus, we concluded that the bridge site (II) is found only in 
the model I with the very thin Pd substrate in the nano-meter scale. 

\section{\label{Reaction energy}Estimation of reaction energy}

To estimate the reaction energy $\Delta E$ of 
NO$_x$ reduction and CO oxidation, 
we consider the next reaction paths. 
The former depends on the species of NO$_x$. 
\[2x({\rm PdO})_4/{\rm Pd}+2{\rm NO}_x
\rightarrow
{\rm N}_2 + 2x ({\rm PdO})_4{\rm O}/{\rm Pd},\]
This process may occur, 
when NO$_x$ molecules dissociate on 
the PdO surface and oxygen atoms stay on the surface. 
However, we do not exclude other complex reaction paths. 
If NO$_x$ molecules dissociate on another catalytic oxide surface, 
and if the adsorbed oxygen atoms move to the PdO surface, 
the process is effectively realized. 
The reaction energy is estimated by the following energy difference. 
\[\Delta E = 
2xE_{\rm PdO} 
+ 2E_{{\rm NO}_x} 
- 2x E_{{\rm PdO}+{\rm O}}
-E_{{\rm N}_2},\]
The latter CO oxidation due to the adsorbed oxygen atoms may be 
given in the next reaction process in the Eley-Rideal reaction 
scheme (the ER scheme).
\[
({\rm PdO})_4{\rm O}/{\rm Pd}+{\rm CO} 
\rightarrow
{\rm CO}_2+({\rm PdO})_4/{\rm Pd}
,\]
The reaction energy is estimated as follows. 
\[\Delta E = 
E_{{\rm PdO}+{\rm O}}
+ E_{{\rm CO}} 
-E_{\rm PdO} 
-E_{{\rm CO}_2}.\]
Here, the energy of a molecule is given by 
a simulation with the same supercell as that utilized 
in simulations of $({\rm PdO})_4/{\rm Pd}$ and 
$({\rm PdO})_4{\rm O}/{\rm Pd}$. 

The reaction energy is summarized 
in Table \ref{Reaction_energy} for NO$_2$, NO, N$_2$O and CO. 
When the value is positive, 
the reaction is exothermic. 
The result suggests that N$_2$O may dissociate on PdO surfaces 
and that the adsorbed oxygen is reactive with CO. 
Although the reduction of NO$_2$ only by oxidation of 
the PdO film is endothermic, 
the absolute value of the reaction energy is 7\,\% of 
the CO oxidation energy for the bridge site (I). 
Considering a successive reaction of NO$_x$ reduction and CO oxidation 
on ($\sqrt{5}\times\sqrt{5}$)$R27^\circ$ PdO, 
we conclude that an exothermic process 
$2{\rm NO}_x + 2x{\rm CO} \rightarrow 
{\rm N}_2 + 2x {\rm CO}_2$ 
is favored at an optimized condition of the temperature and 
the partial pressures of gas components. 

\begin{table}
\begin{center}
\begin{tabular}{|l|l|l|l|l|}
\hline 
Oxygen sites & 
NO$_2$ reduction &
NO reduction &
N$_2$O reduction &
CO oxidation 
\\
&
energy [eV] &
energy [eV] &
energy [eV] &
energy [eV] 
\\
\hline
Bridge (I) &
-0.194 &
0.007 &
0.613 &
2.775 
\\
\hline
Bridge (II) &
-0.403 &
-0.097 &
0.561 &
2.827 
\\
\hline
\end{tabular}
\end{center}
\caption{\label{Reaction_energy}
Reaction energy of NO$_x$ reduction and 
CO oxidation on ($\sqrt{5}\times\sqrt{5}$)$R27^\circ$ PdO/Pd(100).
Definition of each energy is given 
by $\Delta E$ in the text. 
}
\end{table}

To examine possibility on reaction paths,  
values of the adsorption energy of O$_2$ and O were obtained. 
The evaluation is given by 
determining the following energy difference. 

\[O + (PdO)_4/Pd \rightarrow (PdO)_4O/Pd,\]
\[\Delta E = E_{PdO}+ E_O -E_{PdO+O},\]

\[\frac{1}{2}O_2 + (PdO)_4/Pd \rightarrow (PdO)_4O/Pd,\]
\[\Delta E = E_{PdO}+ \frac{1}{2}E_{O_2} -E_{PdO+O},\]

The result shown in Table \ref{Oxygen_ads} suggests that 
O$_2$ molecules in a gas phase do not spontaneously react with 
the PdO film. 
Thus we cannot expect dissociative adsorption of O$_2$ 
on this Pd oxide surface. 
But, if we have atomic oxygen, its reactivity with the 
PdO film is enough high. 
If we have a chance to obtain this active atomic oxygen 
from dissociative adsorption of NO$_x$ somewhere on 
the catalyst surface, the three-way catalytic reaction would work 
smoothly. At the same time, we should consider this oxygen adsorbed PdO film 
as an intermediate structure of the whole catalytic processes. 

\begin{table}
\begin{center}
\begin{tabular}{|l|l|l|}
\hline 
Oxygen sites & 
O adsorption &
O$_2$ adsorption 
\\
&
energy [eV] &
energy [eV] 
\\
\hline
Bridge I &
 2.255 &
-0.069 
\\
\hline
Bridge II &
 2.202 &
-0.121 
\\
\hline
\end{tabular}
\end{center}
\caption{
\label{Oxygen_ads}
Adsorption energy of O and O$_2$ 
on ($\sqrt{5}\times\sqrt{5}$)$R27^\circ$ PdO/Pd(100).
The energy is per the surface unit cell.}
\end{table}

\section{\label{Surface_migration}Surface migration of adsorbed oxygen}

It is well-known that 
the NO$_x$ reduction process consists of 
a complex process with multiple reaction paths.\cite{NO_reaction} 
In our discussion, we assume that 
an oxygen adsorbed PdO/Pd(100) surface 
can be an intermediate state of a total redox reaction. 
To explore all possible processes on the  
oxygen adsorbed surfaces, we need to have 
information on every reaction path and its activation barrier. 
This problem needs rather detailed simulations. 
To test functionality of the 
oxygen adsorbed oxide film as a catalyst, however, 
we can confirm reactivity of the adsorbed oxygen atom. 

\begin{figure}
(a) \\
\begin{center} 
\includegraphics[height=8cm]{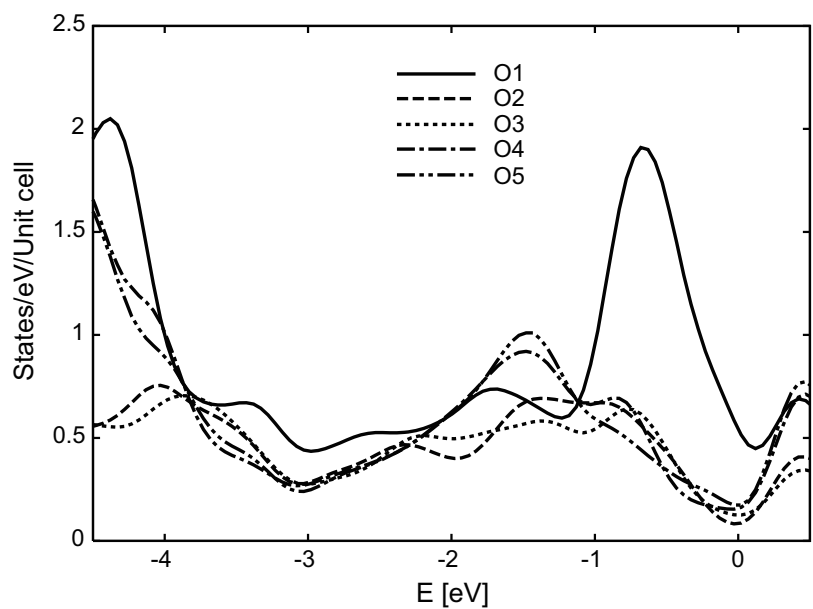} 
\end{center}
\ \\
(b) \\
\begin{center} 
\includegraphics[height=8cm]{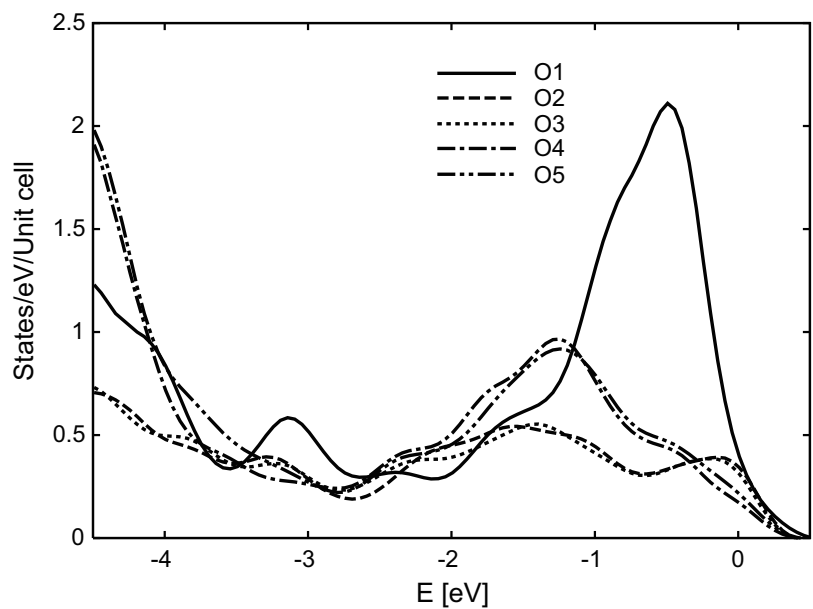}
\end{center}
\caption{
The projected local density of states at oxygen sites 
in the bridge-site (I) structure for 
(a) the model I and (b) the model II. 
The projection is done using oxygen $p$ orbitals. 
The adsorbed oxygen denoted as O1 is at the bridge sites (I). 
Other oxygen atoms (O2 $\cdots$ O5) are those in 
a unit cell of the PdO film. 
The origin of the energy is the Fermi energy. 
\label{LDOS_PdO}}
\end{figure}

A first clue of finite reactivity of the adsorbed oxygen 
is local information of the electronic structure. 
We obtained the local density of states (LDOS) 
given by our GGA simulation. (Figure\,\ref{LDOS_PdO}) 
The data is given by projecting only contribution of 
$p$ orbitals at each oxygen atom. 
Comparison of data on each oxygen atom 
reveals that the monatomic adsorbed oxygen 
at the bridge site (I) 
has an enhancement in LDOS at around the Fermi energy. 
Thus we can conclude that the oxygen has a frontier orbital\cite{Fukui} 
and the reactivity is enhanced at this oxygen site. 

\begin{figure}
\begin{center}
(a)
\includegraphics[height=5cm]{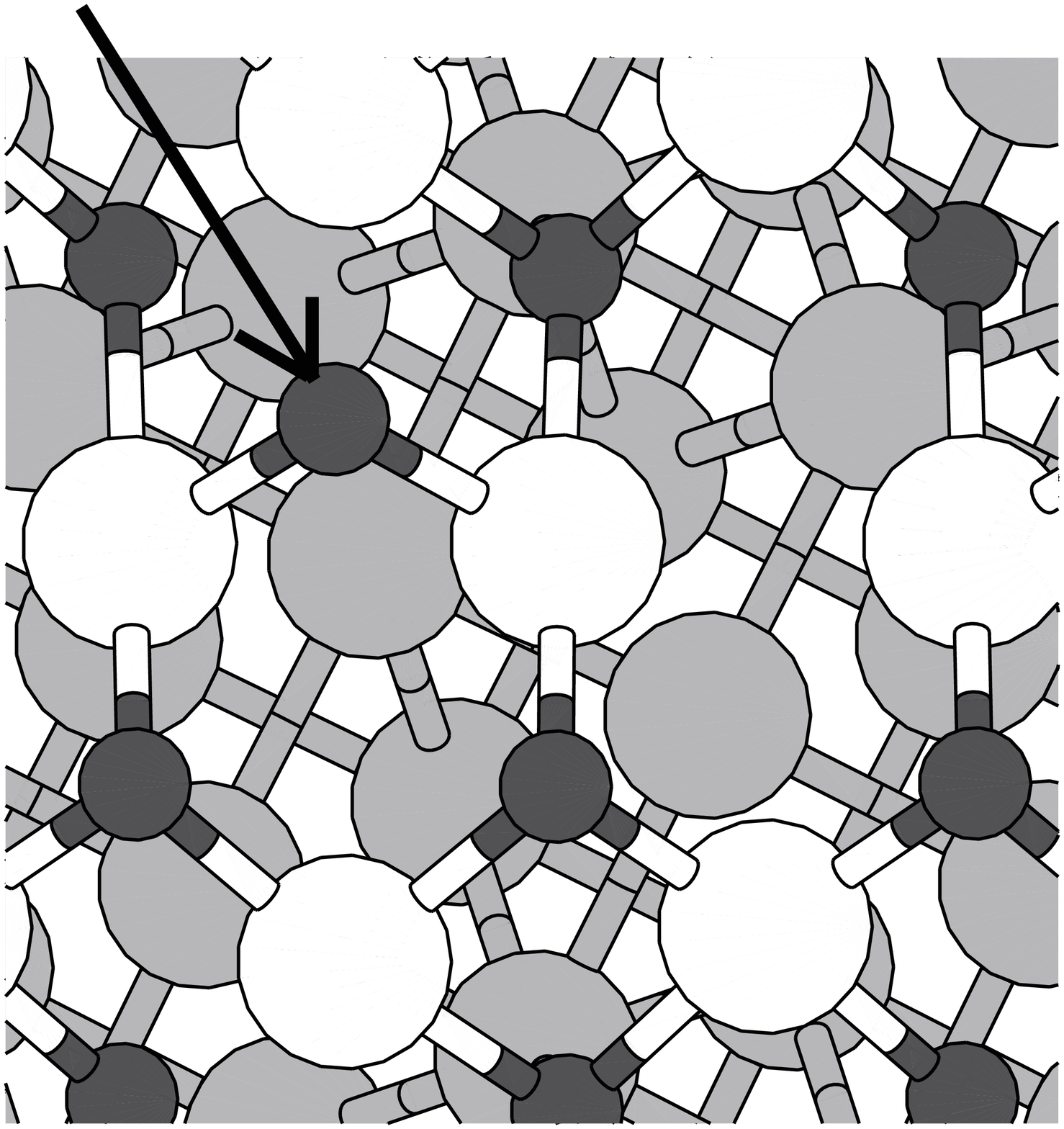}
(b)
\includegraphics[height=5cm]{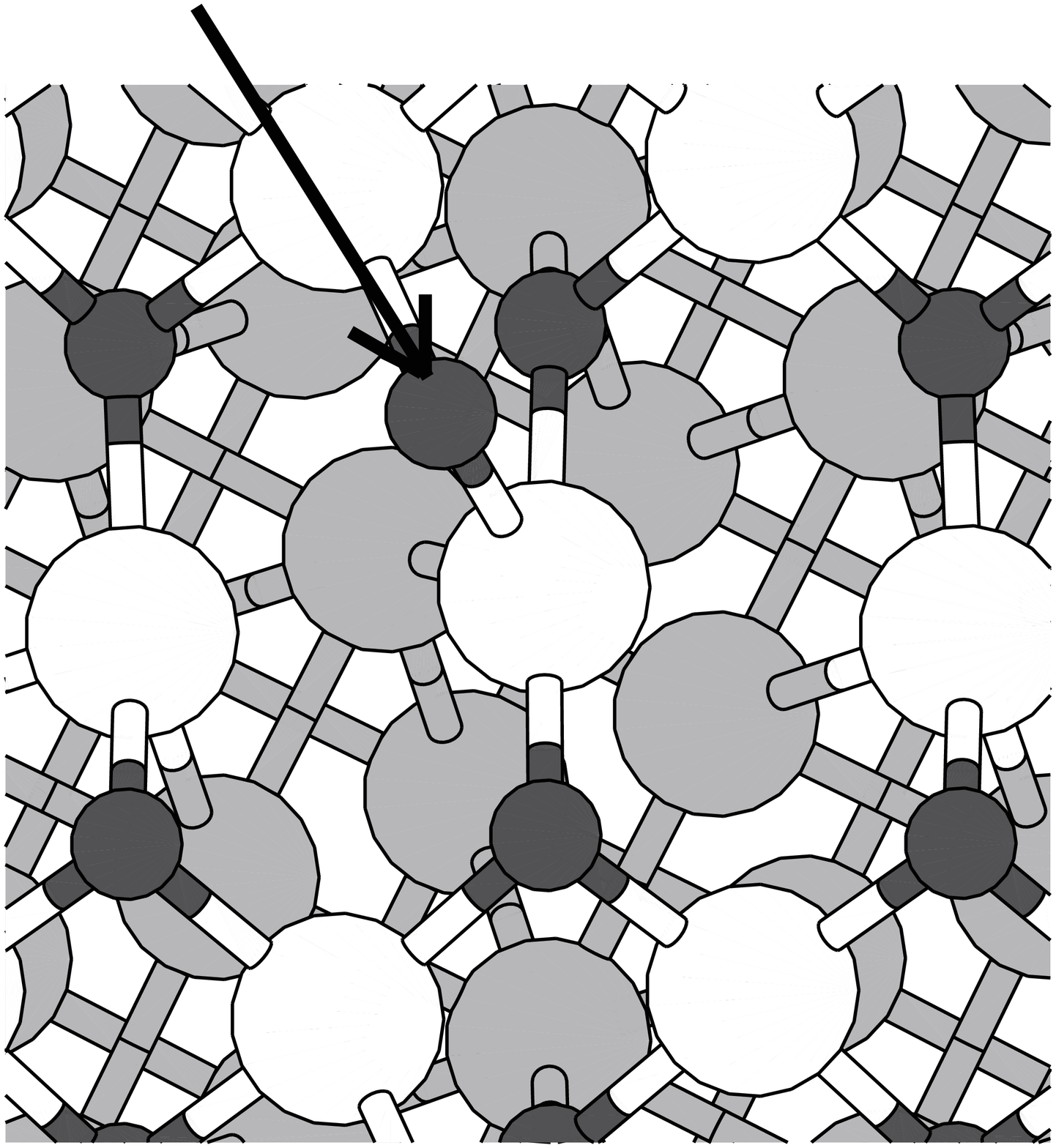}
\ \\
\ \\
(c)
\includegraphics[height=5cm]{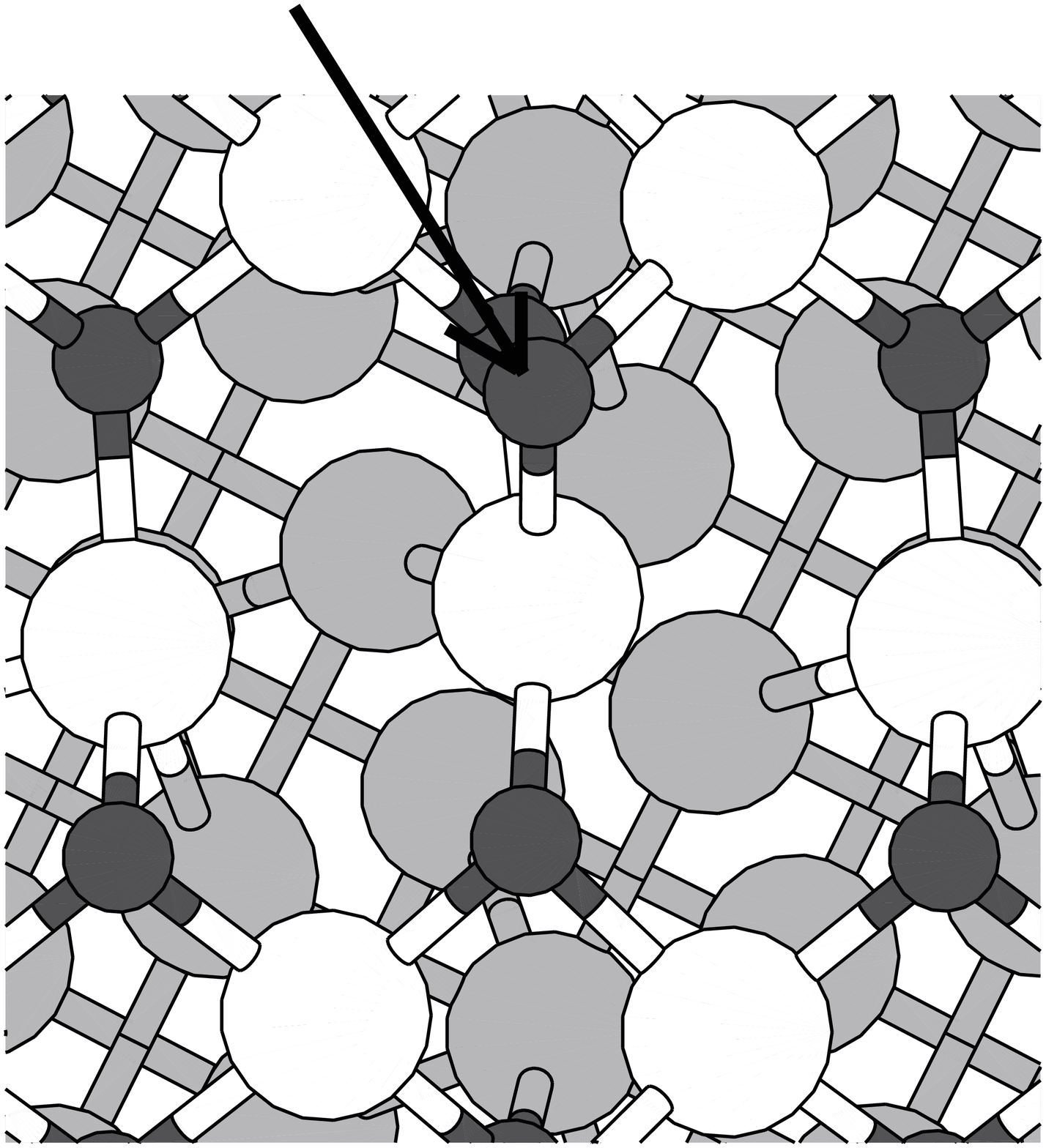}
(d)
\includegraphics[height=5cm]{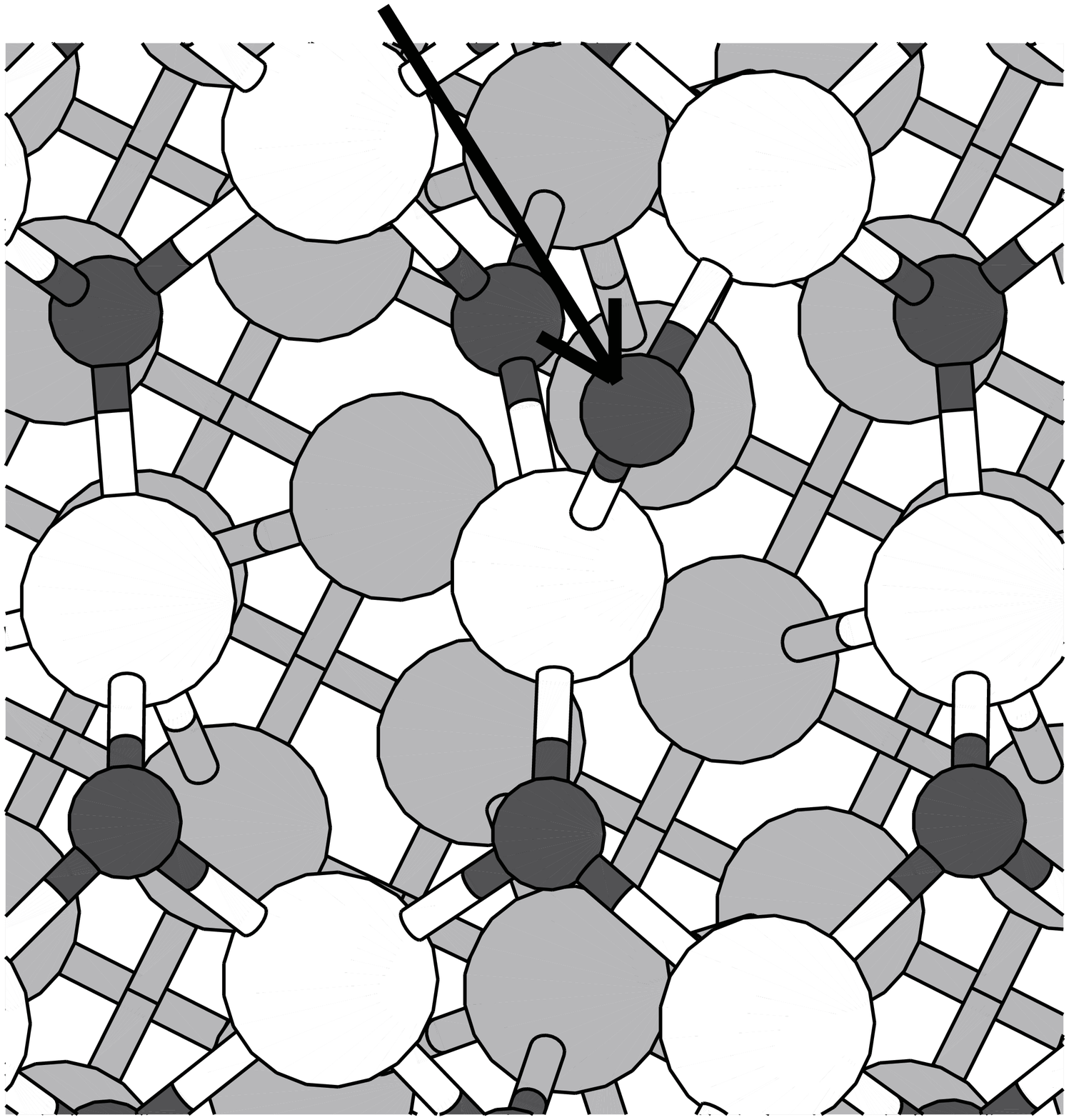}
\caption{
Atomic structure of oxygen adsorbed on the PdO film 
in a transition path from the bridge-site (I) structure 
(a) to the bridge-site (II) structure (d) in the model I. 
The third and the fifth structures in 
totally seven replicas are shown in (b) and (c), respectively. 
The adsorbed oxygen marked by an arrow keeps a bond connection always with 
the 2-fold-coordinated Pd atom at the center in each panel. 
\label{Path_Bridge_Hollow}
}
\end{center}
\end{figure}

This oxygen adatom with high reactivity 
has further mobility on this surface. 
To check possible fluxionality of 
the oxygen adatom, we performed 
estimation of a migration path from 
the bridge site (I) to the bridge site (II) 
for the model I, and 
from the bridge site (I) to the next bridge site (I) 
for the model II. 
The nudged elastic band method (NEBM)\cite{NEB} 
is adapted for this simulation. 
Seven replicas for the model I and 
eight replicas for the model II 
were prepared and the paths were optimized. 
Some snapshots in between these stable sites 
are shown in Figure\,\ref{Path_Bridge_Hollow}. 

The barrier height for this transition of the model I 
is estimated to be $\simeq$ 0.45\,eV. 
Optimization of atomic configuration of the PdO film 
is important to have this value. 
Gradual modification of the PdO structure is seen in 
the panels (b) and (c) of Figure\,\ref{Path_Bridge_Hollow}. 
In this transition, 
the adsorbed oxygen keeps a bond connection with 
a 2-fold-coordinated Pd atom in the PdO film. 
We can interpret that a rotational motion of the extra 
oxygen atom around a O-Pd-O structure 
occurs at the 2-fold-coordinated Pd site in PdO film. 
Thus, the existence of low-coordinated Pd atoms is a key to 
understand possible fluxionality of the adsorbed oxygen. 

\begin{figure}
\begin{center}
\includegraphics[height=8cm]{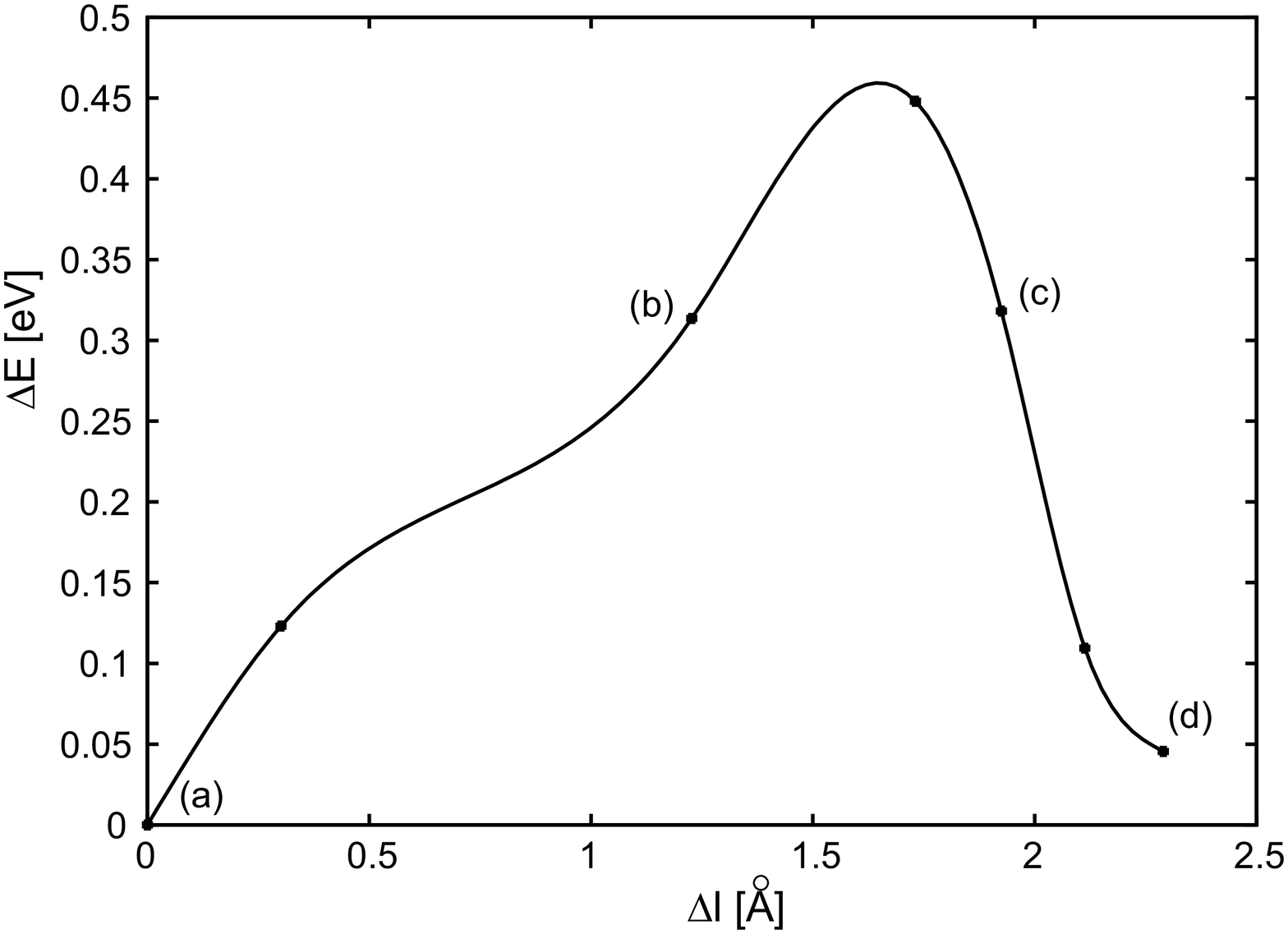}
\caption{Oxygen migration barrier 
from the bridge site (I) to the bridge site (II) in the model I. 
The horizontal axis is given by 
a distance between a 2-fold-coordinated Pd and the active oxygen atom. 
This distance becomes the Pd-O bond length 
of the adsorbed oxygen and a two-fold-coordinated Pd atom in the PdO film. 
The origin of the value, $\Delta l$, 
is shifted by the bond length in the stable bridge site (I). 
The vertical axis represents a migration energy. 
Structures given by Figure 5(b) and (c) are around 
$\Delta l=1.2$ and 1.9, respectively. 
\label{Transition_path}
}
\end{center}
\end{figure}

We determined 
the first barrier of 0.45\,eV 
for the possible total transition 
from a stable bridge-site (I) structure to 
another bridge-site (I) structure in the model I. 
We have another barrier from the bridge site (II) structure 
of Figure\,\ref{Path_Bridge_Hollow}\,(d) to another bridge site (I). 
However, it is much smaller than the above barrier, 
since the next bridge-site (I) structure is easily accessible from 
the bridge-site (II) structure. 
There might be a direct path from a bridge site (I) 
to another bridge site (I), 
but the barrier height should be a little higher than the above value. 
Thus, we assume that a typical barrier height is around 0.45\,eV 
for the migration of oxygen atoms on PdO of the thin model 
(See Figure \ref{Transition_path}). 

A simulation for the model II 
tells us another picture. 
We performed a NEBM simulation 
from the bridge-site (I) structure 
to another neighboring bridge-site (I) structure using the model II. 
In this simulation, the barrier height is found to be $\sim 1.1$eV. 
There appear no bridge-site (II) structure in this process. 
In addition, we found that 
the local structure of the substrate was almost 
kept rigid in the reaction path, although all of atoms were allowed to move. 
This rigidity prevents modification of the PdO film. 
If continual bond formations between the adsorbed oxygen 
and Pd atoms on the path appear, 
the height of the energy barrier decreases. 
But, in this model II, the energy reduction hardly occurs. 
On the contrary, easiness in morphological change in the structure 
allows easy fluxionality of oxygen in the model I. 
Thus, we conclude that the fluxionality may be attributed 
to a nano-meter-scale thin oxidized Pd structure. 

The value of 0.45\,eV for the barrier is rather small. 
If we refer to values in Table \ref{Reaction_energy}, 
we can expect a large kinetic energy of the adsorbed oxygen atoms 
coming from the reaction heat of NO$_x$ reduction. 
The oxygen atoms will move approximately with motional energy 
of typically $\sim$ 1\,eV. 
Thus, the mobility of oxygen atoms would be easily kept in the whole process. 

Here, we should note that known PdO films on various 
Pd surfaces always have 2-fold-coordinated sites 
as well as PdO$_4$ units.\cite{Klikovitz,Westerstrom} 
We can thus hope to have similar migration paths of oxygen 
around the 2-fold-coordinated Pd atoms. 
Here, the bond angle of a structure O-Pd-O shows notable 
difference between the model I and the model II. 
In the model I, the O-Pd-O angles are 135$^\circ$ 
and 160$^\circ$, while 
the values increase and become 
162$^\circ$ and 169$^\circ$ 
in the model II. 
The latter values are approximately equal to 
170$^\circ$ found in the literature.\cite{Kostelnk}
The bond angle itself does not directly reflect the migration barrier. 
However, small angles in O-Pd-O happen 
only for the model I, where relaxation in the PdO film 
structure is easily achieved. 
Thus the value of the migration barrier should reflect 
this structural feature, {\it i.e.} easy structural modification. 
Once the fluxionality of reactive oxygen is guaranteed 
owing to this feature, 
we will have an enhanced cross section for CO oxidization process. 
But, we should note again that the structural feature of 
the model I is lost already for a slab model with 
3 or more Pd layers in the substrate. 

\section{\label{Summary}Summary and Conclusions}

We have obtained two stable structures 
for oxygen-adsorbed PdO film over a Pd(100) thin structure. 
Existence of both 2-fold-coordinated Pd atoms and 4-fold-coordinated Pd atoms 
is important to have fluxionality of oxygen atoms 
at stable bridge sites (I). 
The oxygen atom moves around a 2-fold-coordinated Pd atom, 
which is a characteristic feature of the PdO films. 
On a PdO film formed on a substrate of two Pd layers, 
the oxygen atom may reach at a bridge site (II). 
In this case, the migration barrier may be reduced to 
$\sim 0.45$eV. 
Because half of Pd atoms in the PdO film are 2-fold-coordinated 
and can provide adsorption sites for the activated oxygen atoms, 
the fluxionality of oxygen can happen on the whole PdO film structure. 

Because of enhancement in the reactivity estimated by 
LDOS and fluxionality in migration paths as well as 
the exothermic nature in 
a continual reaction of CO oxidation and N$_x$O reduction, 
we conclude that the continual 
redox reactions may occur through motion and 
reaction of active oxygen atoms on the stable PdO film. 

Now, we have a well-defined picture of catalytic 
reactions of the PdO surfaces with the 2-fold-coordinated Pd atom 
surrounded by two oxygen atoms. 
Here, we should note that 
the local structure of this O-Pd-O bond connections 
is similar to those found in 
Pd catalysts\cite{Hayashi} utilized for 
the C-C cross-coupling reactions of the Suzuki-Miyaura coupling,\cite{Suzuki} 
where ligands might be phosphine or arsine, rather than oxides. 
We can expect that 
the O-Pd-O structure is reactive also for the hydro-carbon structures 
as in the homogenous catalysts. 
Since the redox reactivity for NO$_x$ and CO mixture is energetically 
confirmed as above, 
we can also expect to have a finite reactivity 
for the heterogeneous catalysts 
even in a three-way catalytic reaction. 

When we notice that the above mechanism found in the PdO film 
can be a general one, we have 
an idea to find similar reactive surface in oxide films. 
Oxide films with low coordinated 
metal atoms in an oxide structure 
provide fluxionality of the adsorbed oxygen atoms 
as found in the PdO film. 
Especially, 2-fold-coordinated metal atoms can have important roles. 
If an attached extra oxygen atom rotates around the 2-fold-coordinated metal, 
the oxygen can easily move around on the film by passing from 
a stable oxygen state to the others. 
When a CO molecule in the gas phase collides with the adsorbed 
oxygen atom in the highly reactive state, 
the reaction in the ER scheme happens. 
If the oxygen is in a fluxional state, 
reaction rates are inevitably increased, 
because the cross section of the collision should be enhanced. 
The ER scheme is unlikely to be sensitive to the surface temperature, 
which would be requested for a solid catalyst effective in 
a wide temperature range. 

For a continual reaction, to keep the low-coordination of 
metal atoms in surface oxides is thus the key factor. 
In order to have this functionality, 
next conditions may be requested for a catalytic oxide. 
\begin{enumerate}
\item Existence of fluxionality for oxygen motion 
around a low-coordinated metal atom. 
\item Prevention of highly oxidized state at catalytic sites 
to keep the low coordination. 
\item Prevention of over-reduction at catalytic sites 
to keep the oxide structure. 
\item Existence of a buffer for oxygen atoms in the catalyst 
or in the environment. 
\end{enumerate}
The first point ensures high reactivity of 
the CO reduction process. 
It also relates to reactivity of NO$_x$ reduction, 
since mobility of extra oxygen allows the catalyst also 
to keep a reaction rate of NO$_x$ reduction. 
The second point is required to keep the reaction site active. 
This factor might be much easily achieved, 
if we have another metal species which is much easily oxidized, and 
if the species form a buffer of oxygen atoms inside a catalytic structure. 
The third factor is important to prevent an aging effect, 
where so-called agglomeration of metal particles 
kills the redox functionality. 

To preserve the reactive property, 
the oxygen buffer would be effective. 
The buffer may be a part of the whole catalyst structure, 
{\it i.e.} the substrate. 
It should be less reductive than Pd, since 
oxygen atoms should be kept in the buffer 
under the over-reduction environment. 
For this purpose, a pure metal substrate is less effective. 
Existence of rare-earth elements in an oxide structure, 
for example, would be important for this function. 

Thus realization of the low-coordinated oxide film structures 
even in another bulk oxide structure can be the final answer 
for a highly reactive redox catalyst. 
The oxide support is actually often used in real three-way catalyst. 
If we look at the functionality of the perovskite 
catalyst,\cite{Tanaka,Catalyst} 
the present mechanism may work as a hidden important ingredient 
enhancing the performance. 

\subsection*{Acknowledgement}
The authors thank all of the research members in 
the elements science and technology project entitled 
``New development of self-forming nano-particle catalyst 
without precious metals''. 
One of the author (K.K.) is grateful 
for discussion by Prof. Y. Tobe. 
This work was supported by 
the elements science and technology project 
and also by Grant-in-Aid for Scientific Research in 
Priority Areas (No. 17064006, No. 19051016) and 
a Grants-in-Aid for Scientific Research (No. 19310094). 
The computation is partly done using the computer facility of 
ISSP, Univ. of Tokyo.

\end{document}